\documentclass[%
 reprint,
superscriptaddress,
 amsmath,amssymb,
 aps,
 longbibliography
]{revtex4-1}

\usepackage{graphicx}
\usepackage{dcolumn}
\usepackage{bm}
\usepackage{color}
\usepackage{hyperref}
\usepackage[capitalise]{cleveref}
\usepackage{tikz}
\usetikzlibrary{quantikz}
\crefname{figure}{Fig.}{Fig.}
\usepackage{amsthm}

\newtheorem{lemma}{Lemma}
\theoremstyle{definition}

\theoremstyle{theorem}

\theoremstyle{definition}

\newcommand{\eps}{\epsilon}

\begin{document}

\title{Variational quantum eigensolvers for sparse Hamiltonians}

\author{William M. Kirby}
\affiliation{Department of Physics and Astronomy, Tufts University, Medford, MA 02155}
\email{william.kirby@tufts.edu}

\author{Peter J. Love}
\affiliation{Department of Physics and Astronomy, Tufts University, Medford, MA 02155}
\affiliation{Computational Science Initiative, Brookhaven National Laboratory, Upton, NY 11973}

\begin{abstract}

Hybrid quantum-classical variational algorithms such as the variational quantum eigensolver (VQE) and the quantum approximate optimization algorithm (QAOA) are promising applications for noisy, intermediate-scale quantum (NISQ) computers. Both VQE and QAOA variationally extremize the expectation value of a Hamiltonian.  All work to date on VQE and QAOA has been limited to Pauli representations of Hamiltonians. However, many cases exist in which a sparse representation of the Hamiltonian is known but there is no efficient Pauli representation. We extend VQE to general sparse Hamiltonians. We provide a decomposition of a fermionic second-quantized Hamiltonian into a number of one-sparse, self-inverse, Hermitian terms linear in the number of ladder operator monomials in the second-quantized representation. We provide a decomposition of a general $d$-sparse Hamiltonian into $O(d^2)$ such terms. In both cases a single sample of any term can be obtained using two ansatz state preparations and at most six oracle queries. The number of samples required to estimate the expectation value to precision $\eps$ scales as $\eps^{-2}$ as for Pauli-based VQE. This widens the domain of applicability of VQE to systems whose Hamiltonian and other observables are most efficiently described in terms of sparse matrices. 

\end{abstract}
\maketitle

\section{Introduction}

The leading applications for noisy, intermediate-scale quantum (NISQ) computers are the variational quantum eigensolver (VQE)~\cite{peruzzo14a} and the quantum approximate optimization algorithm (QAOA)~\cite{farhi2014quantum}. VQE estimates the ground state energy (and other properties) of a Hamiltonian by optimizing an ansatz for an energy eigenstate~\cite{peruzzo14a,omalley16a,santagati18a,shen2017quantum,paesani2017,kandala17a,hempel18a,dumitrescu18a,colless18a,nam20a,kokail19a,kandala19a,google20a,kreshchuk2020light}. QAOA approximately optimizes a classical objective function using a parametrized quantum state~\cite{farhi2014quantum}. Methods such as VQE and phase estimation, which compute energy eigenstates, rely on efficient representations of the Hamiltonian. Two such representations are widely used: local Hamiltonians and sparse Hamiltonians.  For simplicity we will restrict our discussion henceforth to systems of qubits; generalizations to tensor factors of arbitrary dimension are straightforward. 

Firstly, we describe the efficient representation of Hamiltonians based on {\em locality}. For this representation, a convenient basis for qubit operators is given by the Pauli operators, $P_i$, which are tensor products of Pauli matrices and the identity. The Hamiltonian is written:
\begin{equation}
    \label{vqehamiltonian_pauli}
    H=\sum_{i=1}^m\alpha_iP_i,
\end{equation}  
where the $\alpha_i$ are real coefficients. This representation of a Hamiltonian is efficient if the number of terms $m$ grows only polynomially with the number of qubits. The locality $k$ of the Hamiltonian \eqref{vqehamiltonian_pauli} is the maximum locality of any term $P_i$, which refers to the number of non-identity tensor factors in $P_i$. Note that locality here does not necessarily refer to geometric locality. For qubits the Hamiltonian is a sum of one-qubit terms, two-qubit terms and so on. This representation was first used for quantum simulation by Lloyd~\cite{lloyd1996universal}, and all VQE algorithms to date make use of this Hamiltonian representation.

A second efficient representation of a Hamiltonian is based on {\em sparsity}, which refers to the maximum number of nonzero entries in any row or column of the Hamiltonian. For example, in a Hamiltonian of the form~\eqref{vqehamiltonian_pauli} the number of nonzero entries in any row or column in the computational basis is bounded above by $m$, so we refer to the Hamiltonian as $m$-sparse. This follows because the Pauli operators are one-sparse: they each have only one nonzero entry in each row and column. However, not every sparse Hamiltonian is local, and many sparse Hamiltonians do not admit a Pauli decomposition~(\ref{vqehamiltonian_pauli}) with a polynomial number of terms. A simple example is the number operator for a bosonic mode encoded as a binary number in qubit computational basis states, which is one-sparse but has an exponential number of Pauli terms.

Quantum simulation of sparse Hamiltonians has undergone extensive study~\cite{aharonov03a,childs03a,berry07a,childs10a,berry12a,berry14a,berry15a,berry15b,low17a,low19a,berry20a}, culminating in algorithms with optimal (for time-independent Hamiltonians)~\cite{low17a,low19a} or near-optimal (for time-dependent Hamiltonians)~\cite{berry20a} scaling with all parameters. In these algorithms a $d$-sparse Hamiltonian is accessed via a pair of oracle unitaries $O_F$ and $O_H$.
$O_F$ returns the location of the $i$th nonzero entry in a given row $x$.
$O_H$ returns the value of the entry in row $x$ and column $y$ to a given precision.
The actions of $O_F$ and $O_H$ are given by
\begin{align}
    &O_F|x,i\rangle=|x,y_i\rangle,\label{O_F_def}\\
    &O_H|x,y\rangle|z\rangle=|x,y\rangle|z\oplus H_{xy}\rangle,\label{O_H_def}
\end{align}
where $x$ is a row-index of $H$ (i.e., a computational basis state), and for $0\leq i\leq d-1$, $y_i$ is column-index of the $i$th nonzero entry in row $x$ of $H$.
The Hamiltonians are given via these oracle unitaries for the sake of modularity: the sparse Hamiltonians are very general, so oracle queries provide a standardized formalism for accessing them.

A VQE algorithm comprises two main components: a quantum subroutine for estimating the expectation value of a Hamiltonian of interest for some parametrized ansatz state, and a classical outer loop that updates the parameters of the ansatz in order to minimize the expected energy~\cite{peruzzo14a}. The quantum subroutine is implemented by separately estimating expectation values of the terms in the Hamiltonian under some decomposition, most commonly the Pauli decomposition \eqref{vqehamiltonian_pauli}.

In this paper, we extend VQE to sparse Hamiltonians, the possibility of which was briefly discussed in the appendix of~\cite{peruzzo14a}. We decompose sparse Hamiltonians into linear combinations of self-inverse one-sparse Hermitian matrices. We then show how to estimate expectation values of these one-sparse terms using two ansatz state preparations and calls to the oracle unitaries defining the Hamiltonian terms. We will show that our algorithm requires at most six oracle queries per measurement circuit.

The class of sparse Hamiltonians that admit description by oracles of the forms \eqref{O_F_def} and \eqref{O_H_def} is much broader than the class of local Hamiltonians, which admit efficient Pauli decompositions and are thus simulable by standard VQE.
To prove that local Hamiltonians are a subset of sparse Hamiltonians with efficient oracle descriptions, it is enough to give oracle descriptions of the Pauli operators, which we do explicitly in \cref{pauli_oracles}.
Hence for any local Hamiltonian, we could first decompose it into Pauli operators~\cite{jordan28a,bravyi02a,seeley12a}, and then simulate each Pauli operator using sparse VQE.
All electronic structure Hamiltonians that can be simulated using standard VQE can be simulated using sparse VQE in this way.
This also provides an example of oracles with simple implementations that are appropriate for NISQ devices.

However, when a Hamiltonian has an efficient Pauli decomposition, it is not a good candidate for sparse VQE because the measurement scheme in \cref{alg} requires an extra qubit and an extra ansatz preparation compared to measuring the Pauli terms directly.
The cases of real interest for sparse VQE are Hamiltonians that are sparse and admit efficient oracle implementations, but do not admit efficient Pauli decompositions.

One such case is a Hamiltonian that includes bosons and is represented in a \emph{direct encoding}~\cite{kreshchuk20a}, in which the occupation of each mode is stored in binary in its own register of qubits.
Bosonic creation and annihilation operators in this encoding are naturally represented in terms of Weyl-Heisenberg shift operators but not as Pauli operators because the occupations of modes can be larger than one.
In \cref{boson_oracles}, we give explicit implementations of oracles for this case.
These implementations can be combined with the oracles for Pauli operators to handle Hamiltonians that act on both fermions and bosons.

The class of sparse Hamiltonians is very large, and we will not attempt to give an exhaustive list of all theories that can be addressed within it.
However, two more examples are as follows.
The first is quantum field theory in \emph{compact encoding}~\cite{kreshchuk20a,kirby2021compactmapping}, in which only the occupations of occupied modes are stored, providing asymptotically optimal space efficiency.
Oracles for field theories in compact encoding are explicitly constructed in~\cite{kirby2021compactmapping}.
The second example is the CI-matrix representation of quantum chemistry, for which oracles are explicitly constructed in~\cite{babbush17a}.

The number of gates and depth of circuits required by the oracles in~\cite{kirby2021compactmapping} and~\cite{babbush17a} are larger than those required for Pauli operators or for quantum field theory in the direct encoding. Implementation of these oracles in the NISQ era would require extensive error mitigation or significantly improved physical gates and qubits. However, sparse VQE will become possible before other sparsity-based simulation algorithms~\cite{berry07a,berry12a,berry14a,berry15a,berry15b,low17a,low19a,berry20a}, because these require more coherent queries to the same oracles.

In \cref{alg} we describe the basic structure of VQE for Hamiltonians that can be decomposed into self-inverse one-sparse Hermitian terms that possess efficient circuit representations. In \cref{decompositions}, we describe methods for obtaining such decompositions. Then in \cref{sec:one_sparse_evolution}, we explain how to construct efficient circuit representations of the resulting terms. These methods permit the implementation of efficient VQE procedures for sparse Hamiltonians. We close the paper with some discussion and directions for future work in \cref{conc}.

\section{Sparse VQE}
\label{alg}

VQE was first used to estimate expectation values of the Hamiltonian~\cite{peruzzo14a}. However, many other quantities are of interest given an ansatz state that is a good approximation to the ground state or other energy eigenstate. For example,~\cite{kreshchuk2020light,kreshchuk20a,kreshchuk2020simulating} study various properties of composite particles in interacting quantum field theory. Properties such as the invariant mass, mass radius, parton distribution function, and form factor are expectation values of corresponding operators, whereas quantities such as the decay constant are matrix elements between different states~\cite{kreshchuk2020simulating}. We will therefore consider estimation of quantities $\langle\phi|\hat O |\psi\rangle$ for sparse operators $\hat O$ between ansatz states $|\phi\rangle = V|0\rangle$ and $|\psi\rangle=U|0\rangle$ prepared by quantum circuits $U$ and $V$. 

We begin with a Hermitian operator that we assume has an efficient decomposition into a sum of Hermitian, self-inverse, one-sparse terms $G_j$:
\begin{equation}
\label{decomp}
\hat O = \sum_{j=1}^t \alpha_j G_j    
\end{equation}
where $\alpha_j$ are real coefficients and the number of terms $t$ is polylogarithmic in the dimension of the Hilbert space on which $\hat O$ acts. If $\hat O$ can be efficiently decomposed into Pauli operators, then the Pauli decomposition of $\hat O$ is an example of~\eqref{decomp} because the Pauli operators are self-inverse and one-sparse. The terms $G_j$ are both Hermitian and unitary, and we further assume that an efficient quantum circuit for each $G_j$ is known. Circuits for sparse unitaries were studied in~\cite{jordan2009efficient}.

Any operator $\hat O$ of the form \eqref{decomp} is sparse, and the number of one-sparse terms $t$ is an upper bound on the sparsity.
In Section~\ref{decompositions} we will explicitly show how to construct a decomposition as in \eqref{decomp} for any arbitrary sparse Hermitian operator. However, for the purpose of this Section it is enough to assume that this is possible, because the actual VQE implementation is agnostic to the method used to obtain the decomposition.

\begin{figure}
    \centering
    \includegraphics[width=0.5\linewidth]{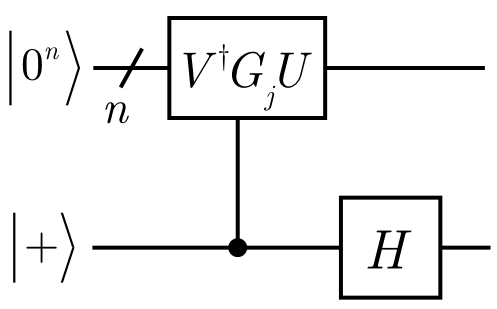}
    \caption{Hadamard test circuit realizing estimation of the real part of the matrix element $\langle0|V^\dagger G_j U|0\rangle$.}
    \label{circfig}
\end{figure}

Given~\eqref{decomp} we perform $M$ Hadamard tests of the operators $V^\dagger G_j U$ via the circuit shown in Figure~\ref{circfig}. This circuit has a state register of $n$ qubits initialized in the all zeros state $|0^n\rangle$ and a single ancilla register initialized in the state $|+\rangle=(|0\rangle+|1\rangle)/\sqrt{2}$. The first operation is application of $V^\dagger G_j U$ controlled on the ancilla qubit. The second operation is a single qubit Hadamard gate applied to the ancilla qubit.
After application of this circuit, the probability of observing zero on the ancilla qubit is:
\begin{equation}\label{prob0}
p(0) = \frac{1}{2}\left(1+{\rm Re}\langle0^n|V^\dagger G_j U|0^n\rangle\right).    
\end{equation}
To replace the real part by the imaginary part of $V^\dagger G_j U$ in~\eqref{prob0}, replace the initial $|+\rangle$ state of the ancilla by the state $|-i\rangle=(|0\rangle-i|1\rangle)/\sqrt{2}$.

After $M$ repetitions of the circuit in Fig.~\ref{circfig}, one obtains $n_0$ zeros and $n_1$ ones from the measurement outcomes of the ancilla qubit. The quantity $(n_0-n_1)/M$ is an estimate of ${\rm Re} \langle0^n|V^\dagger G_j U|0^n\rangle$. We can therefore interpret ancilla outcome $b$ as determining a random variable with value $(-1)^b$. The analysis of the variance of these estimates and hence the scaling of $M$ for given $\hat O$ and precision $\eps$ proceeds exactly as for Pauli decompositions (given in~\cite{mcclean2016theory,rubin2018application}), so the required $M$ for precision $\epsilon$ scales as $\eps^{-2}$.

In this section we have given the extension of VQE to matrices that can be efficiently decomposed into self-inverse one-sparse Hermitian terms described by efficient quantum circuits. Estimation of a matrix element of a self-inverse term $G_j$ between two ansatz states $U|0^n\rangle$ and $V|0^n\rangle$ is accomplished by controlled application of the ansatz circuits $U^\dagger$ and $V$ as well as $G_j$. For estimation of expectation values we have $U=V$ and twice as many ansatz preparations are required as for Pauli decomposition VQE.  The necessity of these extra preparations is apparent when one notes the capability to also estimate matrix elements between distinct states. In the remainder of the paper, we will focus on Hamiltonians for easier comparison to prior literature on sparse Hamiltonian simulation, but all of our results will apply to general sparse, Hermitian observables. We will discuss obtaining the necessary sparse decompositions in Section~\ref{decompositions}, and efficiently applying the resulting operators in Section~\ref{sec:one_sparse_evolution}.

\section{Obtaining sparse decompositions}
\label{decompositions}

Access to a $d$-sparse Hamiltonian is provided by the oracles $O_F$ and $O_H$ as defined in \eqref{O_F_def} and \eqref{O_H_def}.
In the case that $H$ is one-sparse we can simplify the action of $O_F$:
\begin{equation}
\label{one_sparse_enumerator}
    O_F|x,0\rangle=|x,y_x\rangle,
\end{equation}
where $y_x$ is defined to be the column-index of the single nonzero entry in row $x$ (corresponding to $i=0$).

Given a $d$-sparse Hamiltonian, we wish to decompose it into a polynomial number of one-sparse self-inverse Hermitian terms. In some cases of interest, most notably fermionic Hamiltonians in second-quantized form, we already have a decomposition into one-sparse Hermitian terms. A Hamiltonian expressed in second-quantized form is a polynomial of some set of ladder operators for various particles or modes.
The basis for the Hilbert space is given by the occupation number (Fock) representation for each of the modes.
Each term in the Hamiltonian is a monomial of ladder operators, which for fermions in the Fock basis is one-sparse since its action as a linear transformation is to map each single Fock state to some scaling of a single Fock state. Therefore, the fermionic Hamiltonian in the occupation number basis is at most $d$-sparse if it contains $d$ terms, so assuming the number of terms is polynomial in the number of qubits, so is the sparsity.

Ladder operator monomials are in general not self-inverse, nor are they Hermitian. However, for each ladder operator monomial present in the fermionic Hamiltonian, its Hermitian conjugate must also be present, and each such pair together is one-sparse and Hermitian. One-sparseness follows because for a fermionic ladder operator monomial, any given state is mapped to zero by either the monomial or its conjugate (or both); this in fact extends to any Hamiltonian that contains fermionic ladder operators with nonidentity action in every term, even if bosonic operators are also present. To obtain a decomposition into one-sparse terms that are also self-inverse, we use the following Lemma:
\begin{lemma}
\label{decomp_lemma_1}
    Any one-sparse Hamiltonian $H^{(1)}$ may be expressed up to to $L$ bits per real and imaginary part of each entry as a linear combination of
    \begin{equation}
        4L=4\left\lceil\log_2\left(\frac{\sqrt{2}\,\|H^{(1)}\|_\text{max}}{\gamma}\right)\right\rceil
    \end{equation}
    one-sparse, self-inverse Hamiltonians $G_j$, where $\gamma$ is the resulting error in max-norm.
    The $O_F$ oracles for the $G_j$ are the same as the $O_F$ oracle for $H^{(1)}$, and the $O_H$ oracle for any $G_j$ may be computed using two queries to the $O_H$ oracle for $H^{(1)}$.
\end{lemma}
\noindent
The proof may be found in \cref{lemma_1_proof_app}. Note that $\|H^{(1)}\|_\text{max}$ denotes the max-norm of $H^{(1)}$, defined to be the maximum magnitude of any entry in $H^{(1)}$, which is upper-bounded by $\|H^{(1)}\|_\infty$~\cite{childs2010limitations}.

\cref{decomp_lemma_1} is constructive, so we can use the proof to decompose each Hermitian conjugate pair of ladder operator monomials into one-sparse, self-inverse terms. If $N$ is the number of ladder operator monomials in the second-quantized fermionic Hamiltonian, the number of conjugate pairs is at most $N/2$, so \cref{decomp_lemma_1} provides a decomposition of $H$ into a linear combination of at most
\begin{equation}
\label{monomial_decomp}
    2N\left\lceil\log_2\left(\frac{\sqrt{2}\,\|H\|_\text{max}}{\gamma}\right)\right\rceil
\end{equation}
one-sparse, self-inverse Hermitian terms, since the max-norm of each monomial is upper-bounded by $\|H\|_\text{max}$.

Beyond the case of fermionic second-quantized Hamiltonians we consider an arbitrary $d$-sparse Hamiltonian that we only have oracle access to. This includes the case of second-quantized Hamiltonians with both fermionic and bosonic modes. In order to apply \cref{decomp_lemma_1} we first decompose the Hamiltonian into one-sparse terms:
\begin{lemma}[\cite{berry14a}, Lemma 4.4]
\label{decomp_lemma_2}
    If $H$ is a $d$-sparse Hamiltonian, there exists a decomposition $H=\sum_{j=1}^{d^2}H_j$ where each $H_j$ is Hermitian and one-sparse. An $O_F$ query to any $H_j$ can be simulated with two $O_F$ queries to $H$, and an $O_H$ query to any $H_j$ can be simulated with one $O_H$ query to $H$.
\end{lemma}
The proof (in~\cite{berry14a}) is again constructive, so we can use \cref{decomp_lemma_2} to obtain Hermitian one-sparse terms $H_j$, and then use \cref{decomp_lemma_1} to approximately decompose each of these into Hermitian one-sparse, self-inverse terms. The resulting total number of one-sparse, self-inverse terms in the decomposition of $H$ is at most
\begin{equation}
\label{direct_decomp}
    4d^2\left\lceil\log_2\left(\frac{\sqrt{2}\,\|H\|_\text{max}}{\gamma}\right)\right\rceil.
\end{equation}

Comparing \eqref{direct_decomp} and \eqref{monomial_decomp}, we see that in cases where either decomposition could be used, which one is preferable depends on half the number of ladder operator monomials ($N/2$) versus the squared sparsity ($d^2$). For example, the light-front Yukawa model studied in~\cite{kreshchuk20a} leads to a second-quantized Hamiltonian whose sparsity scales as $\Theta(N^{2/3})$.
This is sublinear because each ladder operator monomial maps a large number of Fock states to zero in this model.
However, even though the sparsity is asymptotically smaller than $N$, the squared sparsity is $d^2=\Theta(N^{4/3})$, so for this example it is still better to separately decompose each Hermitian conjugate pair of ladder operator monomials into one-sparse, self-inverse terms using \cref{decomp_lemma_1}.

We use the decomposition provided by \cref{decomp_lemma_2,decomp_lemma_1} because it results in terms that are one-sparse and unitary, and as we will see below, have entries $\pm1$ or $\pm i$; the cost is that the decomposition itself is approximate. However, the resulting terms can be implemented exactly using at most six oracle queries (see \cref{sec:one_sparse_evolution}). Alternative decompositions exist that avoid approximations in the decompositions themselves (e.g.,~\cite{childs11a}), so it is possible that in future the method given above can be improved if the terms in such a decomposition can be implemented using few oracle queries.

\section{Evolution under one-sparse unitary operators}
\label{sec:one_sparse_evolution}

The expectation value estimation method in \cref{alg} requires controlled applications of one-sparse, self-inverse, Hermitian operators; let $G$ be such an operator.
In practice, $G$ will be one of the operators $G_j$ obtained from \cref{decomp_lemma_1}. 
There is an extensive body of methods for simulating sparse Hamiltonians~\cite{aharonov03a,childs03a,berry07a,childs10a,berry12a,berry14a,berry15a,berry15b,low17a,low19a,berry20a}, any of which could be used to implement the controlled application of $G$. However, because $G$ is one-sparse and self-inverse, we can use a simpler method similar to the construction of the quantum walk operator in~\cite{berry12a} (see the proof of Lemma 4 in~\cite{berry12a}). The fact that methods for simulation of time evolution generated by sparse Hamiltonians can also be used for simulation of sparse unitaries was first noted in~\cite{jordan2009efficient}.

Using the oracles $O_F$ and $O_H$ for $G$, we can apply $G$ as follows: let $|x\rangle_s$ be any input computational basis state, and let $|0\rangle_{a_1}|0\rangle_{a_2}$ be ancilla registers. The steps to apply $G$ are: first,
\begin{align}
    |x\rangle_s|0\rangle_{a_1}|0\rangle_{a_2}~&\xrightarrow{O_F}~|x\rangle_s|y_x\rangle_{a_1}|0\rangle_{a_2}\\
    &\xrightarrow{O_H}~|x\rangle_s|y_x\rangle_{a_1}|G_{xy_x}\rangle_{a_2},
\end{align}
where $y_x$ is the column-index of the single nonzero entry in row $x$ of $G$, as in \eqref{one_sparse_enumerator}.
From the proof of \cref{decomp_lemma_1} (in \cref{lemma_1_proof_app}), it follows that $G_{xy_x}=\pm1~\forall x,y_x$ or $G_{xy_x}=\pm i~\forall x,y_x$. Whether the entries are $\pm1$ or $\pm i$ is determined by $j$ (where $G=G_j$ for some $G_j$ resulting from \cref{decomp_lemma_1}), which is evaluated in classical preprocessing.
Therefore, $|G_{xy_x}\rangle_{a_2}$ need only be a single qubit determining the sign, and as our next step we can apply the entry $G_{xy_x}$ exactly as a phase controlled by $|G_{xy_x}\rangle_{a_2}$, and then complete the implementation of $G$ as follows:
\begin{align}
    \xrightarrow{\text{controlled phase}}~&G_{xy_x}|x\rangle_s|y_x\rangle_{a_1}|G_{xy_x}\rangle_{a_2}\label{controlled_phase}\\
    \xrightarrow{O_H^{-1}}~&G_{xy_x}|x\rangle_s|y_x\rangle_{a_1}|0\rangle_{a_2}\\
    \xrightarrow{\text{swap $s,a_1$}}~&G_{xy_x}|y_x\rangle_s|x\rangle_{a_1}|0\rangle_{a_2}\\
    \xrightarrow{O_F^{-1}}~&G_{xy_x}|y_x\rangle_s|0\rangle_{a_1}|0\rangle_{a_2},
\end{align}
where the last step follows because for a one-sparse, Hermitian operator, \eqref{one_sparse_enumerator} implies
\begin{equation}
    O_F|y_x,0\rangle=|y_x,x\rangle.
\end{equation}

The effect of these operations is to map
\begin{equation}
    |x\rangle_s~\mapsto~G_{xy_x}|y_x\rangle_s=G|x\rangle_s,
\end{equation}
i.e., we have applied $G$ to $|x\rangle_s$.
This required four queries to the oracles: one query each to $O_F$, $O_H$, and their inverses.
For ancillas, we required copying the computational register $s$ in the register $a_1$ to apply the $O_F$ oracle, and one additional qubit in the register $a_2$ to represent the sign of $G_{xy_x}=\pm1,\pm i$.
In both queries and ancillas, these are the minimum requirements to apply the oracle unitaries at all.

Finally, recall that the one-sparse, self-inverse terms that we are estimating expectation values of were obtained via \cref{decomp_lemma_1}, above.
However, the full Hamiltonian is first decomposed into one-sparse terms $H^{(1)}$~--- either conjugate pairs of ladder-operator monomials or via \cref{decomp_lemma_2}~--- which form the inputs to \cref{decomp_lemma_1}.
From \cref{decomp_lemma_1} we know that the $O_F$ oracle for any of the $G_j$ is identical to the $O_F$ oracle for $H^{(1)}$.
Also, from the proof of \cref{decomp_lemma_1}, we know that the $O_H$ oracle for any of the $G_j$ can be implemented by first applying the $O_H$ oracle for $G_j$ and then performing a single controlled operation (we would later undo both of these steps to apply $O_H^{-1}$).
Hence the number of queries to each $H^{(1)}$ is still four, each of which will either be implemented directly (in the ladder operator monomial decomposition), or via one (for $O_H$) or two (for $O_F$) queries to the full Hamiltonian using \cref{decomp_lemma_2}.
This gives a total of at most six oracle queries.

\section{Conclusion}\label{conc}

In existing studies, the only Hamiltonian input model used in VQE has been decomposition into Pauli operators. In this paper we have extended VQE to the case of sparse Hamiltonians. We accomplished this by employing a variant of techniques previously considered applicable to future fault-tolerant quantum computers~\cite{berry14a}. For sparse Hamiltonians, we have demonstrated how VQE can be implemented via a decomposition into one-sparse, self-inverse Hermitian terms. As discussed in the introduction, simulation of second-quantized Hamiltonians in condensed matter, high energy and nuclear physics, and in compact representations of quantum chemistry are natural candidates for this sparse VQE method~\cite{aspuru2005simulated,babbush17a,kreshchuk20a,kreshchuk2020light,kreshchuk2020simulating,kirby2021compactmapping}.

This paper focused on VQE, but the results may also be used in the context of QAOA~\cite{farhi2014quantum}. QAOA to date treats classical objective functions that are sums of local clauses, of which $3$-SAT and MAXCUT are canonical NP-complete examples. Classical objective operators are diagonal in the computational basis and hence naturally one-sparse. The techniques here would allow extension to the case where the diagonal entries are given by more complicated classical functions. This broadens the space of examples within which to search for quantum advantage and also may provide practical advantages for problems with large locality such as the travelling salesman problem. We leave the investigation of these ideas to future work.

\begin{acknowledgements}
The authors thank Michael Kreshchuk for helpful conversations and advice. W. M. K. acknowledges support from the National Science Foundation, Grant No. DGE-1842474. This work was supported by the NSF STAQ project (PHY-1818914).
\end{acknowledgements}

\bibliography{biblio.bib}

\appendix

\section[appendix]{Oracle implementations for example applications}

In this section, we give explicit implementations for the oracles used in sparse variational quantum eigensolvers (VQEs), for the special cases of Pauli operators and bosonic creation and annihilation operators.
As described in the main text, a Hamiltonian $H$ can be simulated using sparse VQE when it is sparse and admits description by two oracles $O_F$ and $O_H$, which respectively give the locations and values of the nonzero entries in $H$.
For convenience, we reproduce Eqs.~\eqref{O_F_def} and \eqref{O_H_def} in the main text, which define $O_F$ and $O_H$:
\begin{align}
    &O_F|x,i\rangle=|x,y_i\rangle,\label{O_F_def_app}\\
    &O_H|x,y\rangle|z\rangle=|x,y\rangle|z\oplus H_{xy}\rangle\label{O_H_def_app},
\end{align}
where $x$ is a row-index of $H$ (i.e., a computational basis state), and for $0\leq i\leq d-1$ (where $d$ is the sparsity of $H$), $y_i$ is column-index of the $i$th nonzero entry in row $x$ of $H$.

\subsection[appendix]{Oracles for Pauli operators}
\label[appendix]{pauli_oracles}

We first prove a result characterizing the locations and values of the nonzero entries in arbitrary Pauli operators, then use this to construct oracles for the Pauli operators.

~\\
\noindent
\textbf{Lemma S.1}
\emph{
    For any Pauli operator $P=P_1\otimes P_2\otimes\cdots\otimes P_n$ (for single-qubit Pauli factors $P_k$), let $a$ and $b$ be length-$n$ binary vectors defined bitwise as follows:
     \begin{equation}
     \label{a_def_app}
        a_k\equiv
        \begin{cases}
            0\quad\text{if $P_k=I$ or $Z$},\\
            1\quad\text{if $P_k=X$ or $Y$},
        \end{cases}
    \end{equation}
    and
    \begin{equation}
    \label{b_def_app}
        b_k=
        \begin{cases}
            0\quad\text{if $P_k=I$ or $X$},\\
            1\quad\text{if $P_k=Y$ or $Z$}.
        \end{cases}
    \end{equation}
    Equivalently,
    \begin{equation}
    	P_k=i^{a_kb_k}X^{a_k}Z^{b_k}.
    \end{equation}
    Then for any row index expressed as a binary number $x$, the location $y$ of the nonzero entry in row $x$ of $P$ is
    \begin{equation}
        y=x\oplus a,\label{pauli_loc_app}
    \end{equation}
    where $\oplus$ denotes bitwise XOR.
    The value of the nonzero entry in row $x$ of $P$ is
    \begin{equation}
        p_x=(-1)^{b\cdot x}(-i)^{a\cdot b},\label{pauli_val_app},
    \end{equation}
    where $\cdot$ is the dot product in $\mathbb{Z}_2^n$, e.g.,
    \begin{equation}
        b\cdot x=\sum_{k=1}^{n}b_kx_k.
    \end{equation}
    Note that the binary vectors $a$ and $b$ are the symplectic binary representation of the Pauli operators as in~\cite{gottesman97a}.
}

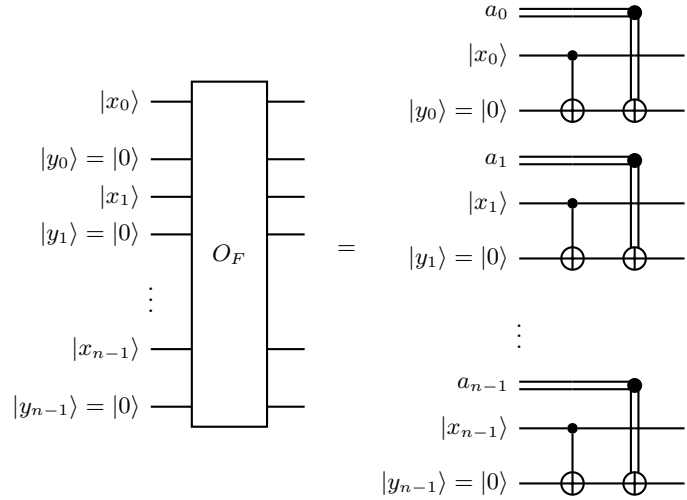
\begin{figure*}
	\centering
	\begin{quantikz}
		\lstick{$\ket{x_0}$} & \gate[wires=7,nwires={5}][1cm]{O_F} & \qw \\
		\lstick{$\ket{y_0}=\ket{0}$} & & \qw \\
		\lstick{$\ket{x_1}$} & & \qw \\
		\lstick{$\ket{y_1}=\ket{0}$} & & \qw \\
		\vdots & \\
		\lstick{$\ket{x_{n-1}}$} & & \qw \\
		\lstick{$\ket{y_{n-1}}=\ket{0}$} & & \qw
	\end{quantikz}
	\quad=
	\begin{quantikz}
		\lstick{$a_0$} & \cw & \cwbend{2} \\
		\lstick{$\ket{x_0}$} & \ctrl{1} & \qw & \qw \\
		\lstick{$\ket{y_0}=\ket{0}$} & \targ{} & \targ{} & \qw \\
		\lstick{$a_1$} & \cw & \cwbend{2} \\
		\lstick{$\ket{x_1}$} & \ctrl{1} & \qw & \qw \\
		\lstick{$\ket{y_1}=\ket{0}$} & \targ{} & \targ{} & \qw \\
		\vdots & \\
		\lstick{$a_{n-1}$} & \cw & \cwbend{2} \\
		\lstick{$\ket{x_{n-1}}$} & \ctrl{1} & \qw & \qw \\
		\lstick{$\ket{y_{n-1}}=\ket{0}$} & \targ{} & \targ{} & \qw
	\end{quantikz}
	\caption{Circuit to implement the $O_F$ oracle as given by \eqref{O_F_Pauli_app}, for the Pauli $P(a,b)$ and input $|x,y=0\rangle$. $x_k$ is the $k$th bit in $x$, $y_k$ is the $k$th bit in $y$, and $a_k$ is the $k$th bit in $a$.}
	\label{pauli_loc_circuit_app}
\end{figure*}

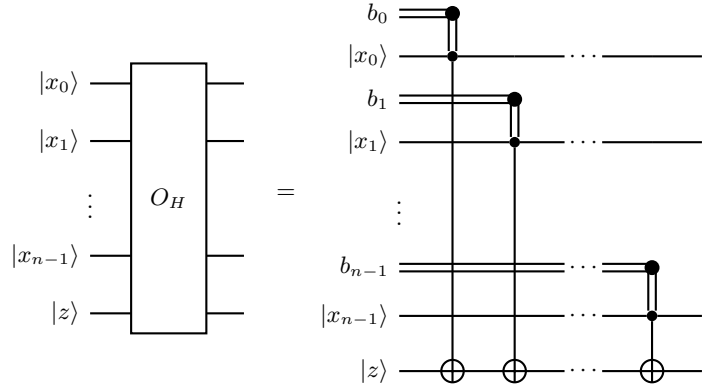
\begin{figure*}
	\centering
	\begin{quantikz}
		\lstick{$\ket{x_0}$} & \gate[wires=5,nwires={3}][1cm]{O_H} & \qw \\
		\lstick{$\ket{x_1}$} & & \qw \\
		\vdots & \\
		\lstick{$\ket{x_{n-1}}$} & & \qw \\
		\lstick{$\ket{z}$} & & \qw
	\end{quantikz}
	\quad=
	\begin{quantikz}
		\lstick{$b_0$} & \cwbend{1} \\
		\lstick{$\ket{x_0}$} & \ctrl{6} & \qw & \qw & \hspace{-0.5cm}\cdots & \qw & \qw \\
		\lstick{$b_1$} & \cw & \cwbend{1} \\
		\lstick{$\ket{x_1}$} & \qw & \ctrl{4} & \qw & \hspace{-0.5cm}\cdots & \qw & \qw \\
		\vdots & \\
		\lstick{$b_{n-1}$} & \cw & \cw & \cw & \hspace{-0.5cm}\cdots & \cwbend{1}  \\
		\lstick{$\ket{x_{n-1}}$} & \qw & \qw & \qw & \hspace{-0.5cm}\cdots & \ctrl{1} & \qw \\
		\lstick{$\ket{z}$} & \targ{} & \targ{} & \qw & \hspace{-0.5cm}\cdots & \targ{} & \qw
	\end{quantikz}
	\caption{Circuit to implement the $O_H$ oracle as given by \eqref{O_H_Pauli_practice_app}, for the Pauli $P(a,b)$ and input $|x\rangle|z\rangle$. $x_k$ is the $k$th bit in $x$, $b_k$ is the $k$th bit in $b$, and $|z\rangle$ is a single qubit. We do not show the register encoding $|x\oplus a\rangle$ as in \eqref{O_H_Pauli_practice_app} because the action of the oracle does not depend on it.}
	\label{pauli_val_circuit_app}
\end{figure*}

\begin{proof}

We first prove \eqref{pauli_loc_app}.
For any computational basis state $|i\rangle$,
\begin{equation}
\begin{split}
	P|i\rangle&=(P_1\otimes P_2\otimes\cdots\otimes P_n)(|i_1\rangle\otimes|i_2\rangle\otimes\cdots\otimes|i_n\rangle)\\
	&=P_1|i_1\rangle\otimes P_2|i_2\rangle\otimes\cdots\otimes P_n|i_n\rangle\\
	&=s|j_1\rangle\otimes|j_2\rangle\otimes\cdots\otimes|j_n\rangle=|j\rangle
\end{split}
\end{equation}
for some $s\in\{\pm1,\pm i\}$.
In other words, each bit $y_k$ in $y$ is equal to the corresponding bit $x_k$ in $x$, acted upon by the corresponding single-qubit Pauli operator $P_k$ in the tensor factorization of $P$.
If $P_k$ is $X$ or $Y$, then the bit is flipped, i.e., $y_k=x_k\oplus1$; otherwise, $y_k=x_k$.
Hence,
\begin{equation}
\label{bitwise_pauli_loc_app}
	y_k=x_k\oplus a_k,
\end{equation}
where $a_k=1$ if $P_k$ is $X$ or $Y$ and $a_k=0$ otherwise.
But this means that when we combine the bits $a_k$ together into the binary vector $a$, we have exactly the definition in \eqref{a_def_app}, and our desired result \eqref{pauli_val_app} simply summarizes the bitwise relations \eqref{bitwise_pauli_loc_app} for all $k$.

We now prove \eqref{pauli_val_app}.
First, note that $P$ can be expressed as
\begin{equation}
\label{pauli_split_app}
	P=s(P^{(Z)}_1\otimes P^{(Z)}_2\otimes\cdots\otimes P^{(Z)}_n)(P^{(X)}_1\otimes P^{(X)}_2\otimes\cdots\otimes P^{(X)}_n)
\end{equation}
for some $s\in\{\pm1,\pm i\}$, where each single-qubit Pauli matrix $P^{(X)}_k$ is either $I$ or $X$, each single qubit Pauli matrix $P^{(Z)}_k$ is either $I$ or $Z$.
The value of $s$ is determined by the fact that $ZX=iY$, so for each $k$ such that $P^{(X)}_k=X$ and $P^{(Z)}_k=Z$ we pick up a factor of $i$ that needs to be cancelled on the right-hand side of \eqref{pauli_split_app}.
The number of these factors is therefore $a\cdot b$, so $s=(-i)^{a\cdot b}$ to cancel them.

The action of $P$ on a computational basis state $|x\rangle$ is
\begin{equation}
\label{sign_action_app}
\begin{split}
	P|x\rangle&=(-i)^{a\cdot b}(P^{(Z)}_1\otimes P^{(Z)}_2\otimes\cdots\otimes P^{(Z)}_n)|y\rangle\\
	&=(-i)^{a\cdot b}P^{(Z)}_1|y_1\rangle\otimes P^{(Z)}_2|y_2\rangle\otimes\cdots\otimes P^{(Z)}_n|y_n\rangle.
\end{split}
\end{equation}
In this expression, the tensor factor for the $k$th qubit is
\begin{equation}
\begin{split}
	P^{(Z)}_k|y_k\rangle&=
	\begin{cases}
		|y_k\rangle\quad&\text{if $P^{(Z)}_k=I$},\\
		(-1)^{y_k}|y_k\rangle\quad&\text{if $P^{(Z)}_k=Z$}
	\end{cases}\\
	&=
	\begin{cases}
		|y_k\rangle\quad&\text{if $b_k=0$},\\
		(-1)^{y_k}|y_k\rangle\quad&\text{if $b_k=1$}
	\end{cases}\\
	&=(-1)^{b_ky_k}|j_k\rangle.
\end{split}
\end{equation}
We can plug this back into \eqref{sign_action_app} to obtain
\begin{equation}
	P|x\rangle=s\prod_{k=1}^n(-1)^{b_ky_k}|y\rangle=(-1)^{b\cdot y}(-i)^{a\cdot b}|y\rangle.
\end{equation}
Hence
\begin{equation}
	(-1)^{b\cdot y}(-i)^{a\cdot b}=(-1)^{b\cdot(a\oplus x)}(-i)^{a\cdot b}
\end{equation}
is the nonzero value in the $x$th column of $P$.
We want the value in the $x$th row, but by \eqref{pauli_loc_app} transposing $P$ simply corresponds to mapping $x$ to $x\oplus a$, so the nonzero value in the $x$th row of $P$ is
\begin{equation}
	(-1)^{b\cdot(x\oplus a\oplus a)}(-i)^{a\cdot b}=(-1)^{b\cdot x}(-i)^{a\cdot b},
\end{equation}
as desired.

\end{proof}

Using Lemma S.1, the oracles for a Pauli operator $P(a,b)$ are defined as follows:
\begin{align}
    &O_F|x\rangle=|x,x\oplus a\rangle,\label{O_F_Pauli_app}\\
    &O_H|x,y\rangle|z\rangle=|x,y\rangle|z\oplus H_{xy}\rangle\label{O_H_Pauli_app}.
\end{align}
Note that there is no argument $i$ in \eqref{O_F_Pauli_app}, unlike \eqref{O_F_def_app}: this is because there is only one nonzero entry per row in a Pauli operator, so there is no need to index the nonzero entries.
The binary vectors $a$ and $b$ are defined in terms of $P(a,b)$ as in \eqref{a_def_app} and \eqref{b_def_app}, respectively, and
\begin{equation}
	H_{xy}=
	\begin{cases}
		(-1)^{b\cdot x}(-i)^{a\cdot b}\quad&\text{if $y=x\oplus a$},\\
		0\quad&\text{otherwise}.
	\end{cases}
\end{equation}
Recall that for sparse VQE, one only needs to evaluate $O_H$ for states that are the output of $O_F$.
Also, the factor of $(-i)^{a\cdot b}$ in the first case above is an overall phase that we can compute classically, so to encode the matrix element it is enough to compute $(-1)^{b\cdot x}$ on the quantum computer.
Hence, in fact we will only ever need to compute
\begin{equation}
	O_H|x,x\oplus a\rangle|z\rangle=|x,x\oplus a\rangle\left|z\oplus(-1)^{b\cdot x}\right\rangle\label{O_H_Pauli_practice_app},
\end{equation}
where $|z\rangle$ is a single qubit that encodes $\pm1$ as
\begin{equation}
\begin{split}
	+1&\mapsto|0\rangle,\\
	-1&\mapsto|1\rangle.
\end{split}
\end{equation}

$O_F$ as given by \eqref{O_F_Pauli_app} is implemented via a single layer of CNOTs and a single layer of classically controlled NOTs, as shown in \cref{pauli_loc_circuit_app}.
$O_H$ as given by \eqref{O_H_Pauli_practice_app} is implemented via a sequence of CNOTs (which are also classically controlled), as shown in \cref{pauli_val_circuit_app}.

\subsection[appendix]{Oracles for Bosonic operators}
\label[appendix]{boson_oracles}

\begin{figure*}
	\centering
	\begin{quantikz}
		\lstick{$v_0$} & \cw & \cwbend{2}\gategroup[wires=15,steps=1,style={dashed,rounded corners,fill=blue!20, inner xsep=2pt},background]{{adds}} & \cwbend{2}\gategroup[wires=6,steps=1,style={dashed,rounded corners,fill=blue!20, inner xsep=2pt},background]{{carry}} \\
		\lstick{$\ket{w_0}$} & \ctrl{1} & \qw & \qw & \qw & \qw & \qw & \qw & \qw & \qw & \qw & \qw & \qw & \qw & \qw & \qw & \qw & \qw & \qw & \qw & \qw \\
		\lstick{$\ket{w'_0}=\ket{0}$} & \targ{} & \targ{} & \octrl{3} & \qw & \qw & \qw & \qw & \qw & \qw & \qw & \qw & \qw & \qw & \qw & \qw & \qw & \qw & \qw & \qw & \qw \\
		\lstick{$v_1$} & \cw & \cwbend{2} & \cw & \cw\gategroup[wires=6,steps=4,style={dashed,rounded corners,fill=blue!20, inner xsep=2pt},background]{{carry}} & \cw & \cwbend{2} \\
		\lstick{$\ket{w_1}$} & \ctrl{1} & \qw & \qw & \ctrl{1} & \ctrl{1} & \qw & \ctrl{1} & \qw & \qw & \qw & \qw & \qw & \qw & \qw & \qw & \qw & \qw & \qw & \qw & \qw \\
		\lstick{$\ket{w'_1}=\ket{0}$} & \targ{} & \targ{}& \targ{} & \octrl{3} & \targ{} & \octrl{3} & \targ{} & \qw & \qw & \qw & \qw & \qw & \qw & \qw & \qw & \qw & \qw & \qw & \qw & \qw \\
		\lstick{$v_2$} & \cw & \cwbend{2} & \cw & \cw & \cw & \cw & \cw & \cw\gategroup[wires=6,steps=4,style={dashed,rounded corners,fill=blue!20, inner xsep=2pt},background]{{carry}} & \cw & \cwbend{2} \\
		\lstick{$\ket{w_2}$} & \ctrl{1} & \qw & \qw & \qw & \qw & \qw & \qw & \ctrl{1} & \ctrl{1} & \qw & \ctrl{1} & \qw & \qw & \qw & \qw & \qw & \qw & \qw & \qw & \qw \\
		\lstick{$\ket{w'_2}=\ket{0}$} & \targ{} & \targ{} & \qw & \targ{} & \qw & \targ{} & \qw & \octrl{3} & \targ{} & \octrl{3} & \targ{} & \qw & \qw & \qw & \qw & \qw & \qw & \qw & \qw & \qw \\
		\lstick{$v_3$} & \cw & \cwbend{2} & \cw & \cw & \cw & \cw & \cw & \cw & \cw & \cw & \cw & \cw\gategroup[wires=6,steps=4,style={dashed,rounded corners,fill=blue!20, inner xsep=2pt},background]{{carry}} & \cw & \cwbend{2} \\
		\lstick{$\ket{w_3}$} & \ctrl{1} & \qw & \qw & \qw & \qw & \qw & \qw & \qw & \qw & \qw & \qw & \ctrl{1} & \ctrl{1} & \qw & \ctrl{1} & \qw & \qw & \qw & \qw & \qw \\
		\lstick{$\ket{w'_3}=\ket{0}$} & \targ{} & \targ{} & \qw & \qw & \qw & \qw & \qw & \targ{} & \qw & \targ{} & \qw & \octrl{3} & \targ{} & \octrl{3} & \targ{} & \qw & \qw & \qw & \qw & \qw \\
		\lstick{$v_4$} & \cw & \cwbend{2} & \cw & \cw & \cw & \cw & \cw & \cw & \cw & \cw & \cw & \cw & \cw & \cw & \cw & \cw\gategroup[wires=4,steps=4,style={dashed,rounded corners,fill=blue!20, inner xsep=2pt},background]{{carry}} & \cw & \cwbend{2} \\
		\lstick{$\ket{w_4}$} & \ctrl{1} & \qw & \qw & \qw & \qw & \qw & \qw & \qw & \qw & \qw & \qw & \qw & \qw & \qw & \qw & \ctrl{1} & \ctrl{1} & \qw & \ctrl{1} & \qw \\
		\lstick{$\ket{w'_4}=\ket{0}$} & \targ{} & \targ{} & \qw & \qw & \qw & \qw  & \qw & \qw & \qw & \qw & \qw & \targ{} & \qw & \targ{} & \qw & \octrl{1} & \targ{} & \octrl{1} & \targ{} & \qw \\
		\lstick{multiplier~$=\ket{0}$} & \qw & \qw & \qw & \qw & \qw & \qw  & \qw & \qw & \qw & \qw & \qw & \qw & \qw & \qw & \qw & \targ{} & \qw & \targ{} & \qw & \qw
	\end{quantikz}
	\caption{Circuit to implement $|w\rangle|w'=0\rangle\mapsto|w\rangle|w'=w\oplus v\rangle$, for a classical binary number $v$ and a binary number $w$ stored in qubits. $w$ represents the occupation of a mode in the input state $x$; $w_{N-1}$ is the most significant bit and $w_0$ is the least. $w'$ represents the occupation of the corresponding mode in the output state, which is initially set to $|0\rangle$. The circuit shown is for five bits in $w$ and $w'$; to generalize to $N$ bits, one should extend the `adds' column and continue to cascade the `carry' blocks as in the example above. If there is a carry from the most significant bit, it means that $w'$ has reached or exceeded the occupation cutoff. Therefore, we flip the `multiplier' qubit from $|1\rangle$ to $|0\rangle$ if there is a carry from the most significant bit: the matrix element will later be multiplied by the value of the `multiplier' qubit, so this sets the matrix element to zero when the occupation reaches the cutoff.}
	\label{increment_circuit}
\end{figure*}
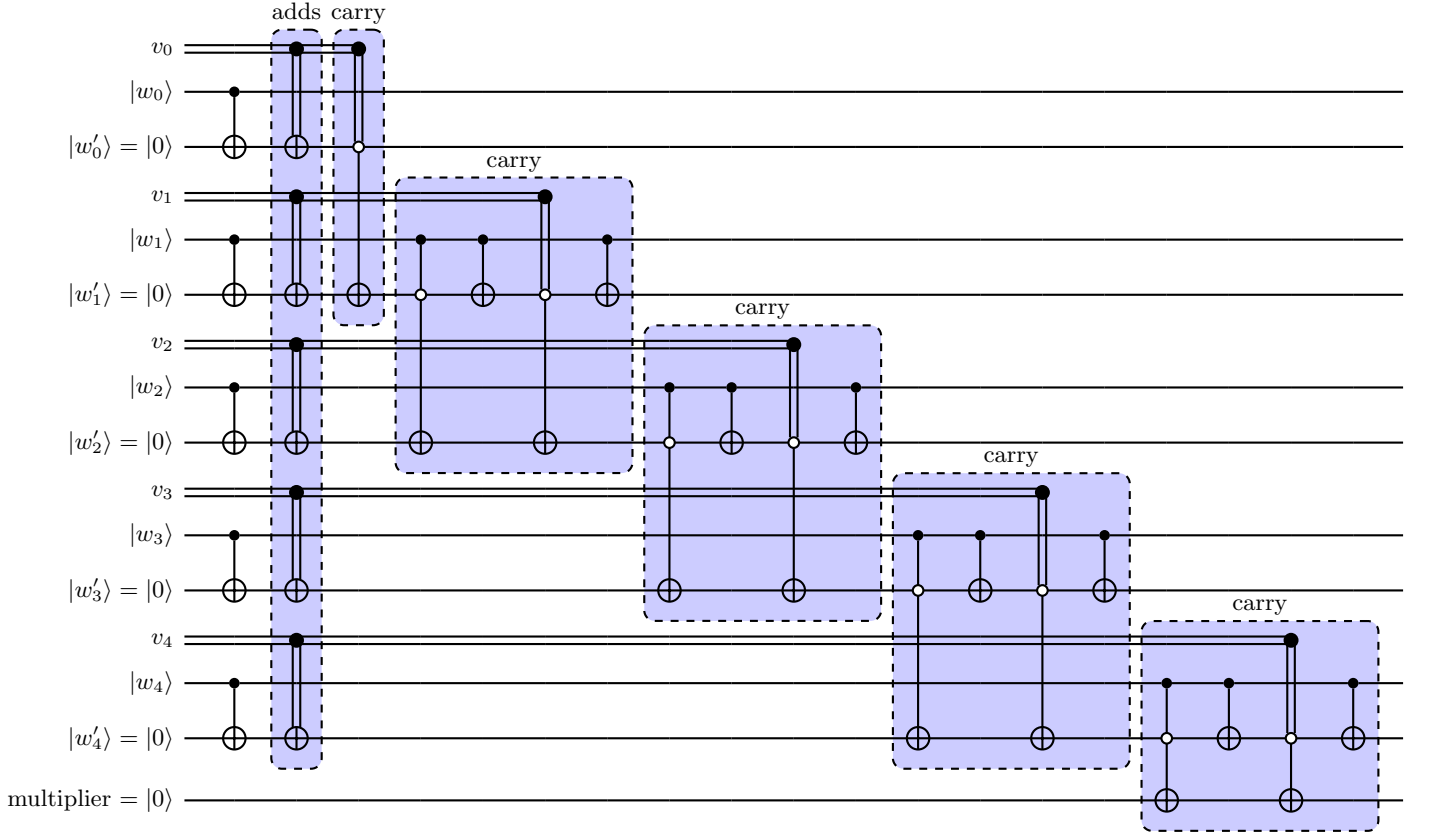

\begin{figure*}
	\centering
	\begin{quantikz}
		\lstick{$v_0$} & \cw & \cwbend{2}\gategroup[wires=15,steps=1,style={dashed,rounded corners,fill=blue!20, inner xsep=2pt},background]{{adds}} & \cwbend{2}\gategroup[wires=6,steps=1,style={dashed,rounded corners,fill=blue!20, inner xsep=2pt},background]{{carry}} \\
		\lstick{$\ket{w_0}$} & \ctrl{1} & \qw & \qw & \qw & \qw & \qw & \qw & \qw & \qw & \qw & \qw & \qw & \qw & \qw & \qw & \qw & \qw & \qw & \qw & \qw \\
		\lstick{$\ket{w'_0}=\ket{0}$} & \targ{} & \targ{} & \ctrl{3} & \qw & \qw & \qw & \qw & \qw & \qw & \qw & \qw & \qw & \qw & \qw & \qw & \qw & \qw & \qw & \qw & \qw \\
		\lstick{$v_1$} & \cw & \cwbend{2} & \cw & \cw\gategroup[wires=6,steps=4,style={dashed,rounded corners,fill=blue!20, inner xsep=2pt},background]{{carry}} & \cw & \cwbend{2} \\
		\lstick{$\ket{w_1}$} & \ctrl{1} & \qw & \qw & \octrl{1} & \ctrl{1} & \qw & \ctrl{1} & \qw & \qw & \qw & \qw & \qw & \qw & \qw & \qw & \qw & \qw & \qw & \qw & \qw \\
		\lstick{$\ket{w'_1}=\ket{0}$} & \targ{} & \targ{}& \targ{} & \ctrl{3} & \targ{} & \octrl{3} & \targ{} & \qw & \qw & \qw & \qw & \qw & \qw & \qw & \qw & \qw & \qw & \qw & \qw & \qw \\
		\lstick{$v_2$} & \cw & \cwbend{2} & \cw & \cw & \cw & \cw & \cw & \cw\gategroup[wires=6,steps=4,style={dashed,rounded corners,fill=blue!20, inner xsep=2pt},background]{{carry}} & \cw & \cwbend{2} \\
		\lstick{$\ket{w_2}$} & \ctrl{1} & \qw & \qw & \qw & \qw & \qw & \qw & \octrl{1} & \ctrl{1} & \qw & \ctrl{1} & \qw & \qw & \qw & \qw & \qw & \qw & \qw & \qw & \qw \\
		\lstick{$\ket{w'_2}=\ket{0}$} & \targ{} & \targ{} & \qw & \targ{} & \qw & \targ{} & \qw & \ctrl{3} & \targ{} & \octrl{3} & \targ{} & \qw & \qw & \qw & \qw & \qw & \qw & \qw & \qw & \qw \\
		\lstick{$v_3$} & \cw & \cwbend{2} & \cw & \cw & \cw & \cw & \cw & \cw & \cw & \cw & \cw & \cw\gategroup[wires=6,steps=4,style={dashed,rounded corners,fill=blue!20, inner xsep=2pt},background]{{carry}} & \cw & \cwbend{2} \\
		\lstick{$\ket{w_3}$} & \ctrl{1} & \qw & \qw & \qw & \qw & \qw & \qw & \qw & \qw & \qw & \qw & \octrl{1} & \ctrl{1} & \qw & \ctrl{1} & \qw & \qw & \qw & \qw & \qw \\
		\lstick{$\ket{w'_3}=\ket{0}$} & \targ{} & \targ{} & \qw & \qw & \qw & \qw & \qw & \targ{} & \qw & \targ{} & \qw & \ctrl{3} & \targ{} & \octrl{3} & \targ{} & \qw & \qw & \qw & \qw & \qw \\
		\lstick{$v_4$} & \cw & \cwbend{2} & \cw & \cw & \cw & \cw & \cw & \cw & \cw & \cw & \cw & \cw & \cw & \cw & \cw & \cw\gategroup[wires=4,steps=4,style={dashed,rounded corners,fill=blue!20, inner xsep=2pt},background]{{carry}} & \cw & \cwbend{2} \\
		\lstick{$\ket{w_4}$} & \ctrl{1} & \qw & \qw & \qw & \qw & \qw & \qw & \qw & \qw & \qw & \qw & \qw & \qw & \qw & \qw & \octrl{1} & \ctrl{1} & \qw & \ctrl{1} & \qw \\
		\lstick{$\ket{w'_4}=\ket{0}$} & \targ{} & \targ{} & \qw & \qw & \qw & \qw  & \qw & \qw & \qw & \qw & \qw & \targ{} & \qw & \targ{} & \qw & \ctrl{1} & \targ{} & \octrl{1} & \targ{} & \qw \\
		\lstick{multiplier~$=\ket{0}$} & \qw & \qw & \qw & \qw & \qw & \qw  & \qw & \qw & \qw & \qw & \qw & \qw & \qw & \qw & \qw & \targ{} & \qw & \targ{} & \qw & \qw
	\end{quantikz}
	\caption{Circuit to implement $|w\rangle|w'=0\rangle\mapsto|w\rangle|w'=w\ominus v\rangle$, for a classical binary number $v$ and a binary number $w$ stored in qubits. $w$ represents the occupation of a mode in the input state $x$; $w_{N-1}$ is the most significant bit and $w_0$ is the least. $w'$ represents the occupation of the corresponding mode in the output state, which is initially set to $|0\rangle$. The circuit shown is for five bits in $w$ and $w'$; to generalize to $N$ bits, one should extend the `adds' column and continue to cascade the `carry' blocks as in the example above. If there is a carry from the most significant bit, it means that $w'$ has reached or exceeded the occupation cutoff. Therefore, we flip the `multiplier' qubit from $|1\rangle$ to $|0\rangle$ if there is a carry from the most significant bit: the matrix element will later be multiplied by the value of the multiplier qubit, so this sets the matrix element to zero when the occupation reaches the cutoff.}
	\label{decrement_circuit}
\end{figure*}

As our second example of oracle implementations, we consider a Hamiltonian that acts on $M$ bosonic modes, each of which is encoded in $N$ qubits.
For simplicity, assume that the cutoff $\Lambda$ on the occupation of each mode is a power of two, so $N=\log_2\Lambda$.
Assume that the bosonic Hamiltonian is written as a polynomial in the creation and annihilation operators $a_p^\dagger$ and $a_p$, where $p$ indexes the mode acted upon.
We want to give oracles for each monomial in the Hamiltonian, since in that case we can use sparse VQE to estimate their expectation values, and hence the expectation value of the Hamiltonian.

These oracles will turn out to involve only binary arithmetic operations on the mode occupation encodings.
Reversible arithmetic is an important subroutine in many quantum algorithms, so various compilers have been designed specifically for this purpose, see for example~\cite{JavadiAbhari2014scaffcc}.
However, for completeness we will give details below of how to implement the specific operations required for the present case.

First consider the location oracle $O_F$.
A monomial of creation and annihilation operators is one-sparse in the Fock (occupation number) basis, so given any input Fock state (encoded in qubits as above), we just need to identify the single outgoing Fock state under the action of the monomial.
Assuming the monomial is normal ordered, we first have a sequence of annihilation operators, then a sequence of creation operators.
For any one of these operators, the mode it acts upon is specified classically, so we just need to either decrement (for annihilation operators) or increment (for creation operators) the occupancy $w$ of the mode.
The occupancies of the modes are encoded as binary numbers, so these operations are just binary addition or subtraction of the number of copies of the operator acting on the mode.
For a given monomial, we determine classically what the corresponding additions or subtractions for each mode are.
On the quantum computer, then, we just need to implement these classically-controlled circuits. 
Circuits to implement these operations are given in \cref{increment_circuit,decrement_circuit}.

We now describe the action of the increment circuit \cref{increment_circuit} on computational basis states: the action on generic states is obtained by linearity.
The first column of CNOTs copies (in the computational basis) the input occupation $|w\rangle$ to the register $|w'\rangle$ that will end up storing the output occupation.
The column labeled `adds' then adds each bit of the classical number $v$ to the corresponding bit in $w'$.
The first `carry' block then adds the bit carried from the zeroth place to $w'_1$: the carried bit from the zeroth place is $1$ if and only if $v_0=1$ and $w'_0=0$ (in which case $w_0$ must be $1$).
Subsequent `carry' blocks add the bit carried from each place to the next more significant place.
Within each `carry' block (after the first), there are two components.
For a given place $k$, if $w_k=1$ and $w'_k=0$, then either $v_k=1$ or there was a $1$ carried from the previous place (but not both), and either way there is a $1$ carried to the next place: this is the first multiply-controlled-NOT in the `carry' block.
If $w'_k=w_k$ and $v_k=1$, then there must have been a $1$ carried from the previous place, so there is a $1$ carried to the next place: the remaining three gates in the `carry' block implement this operation.
However, if $w'_k=w_k$ and $v_k=0$, then there cannot have been a $1$ carried from the previous place, so there is no $1$ carried to the next place.
Similarly, if $w_k=0$ and $w'_k=1$, then either $v_k=1$ or there was a $1$ carried from the previous place (but not both), and either way there is no $1$ carried to the next place.
The multiply-controlled-NOTs in each `carry' block are mutually exclusive, so at most one bit is carried to the next place.
The logic in the decrement circuit \cref{decrement_circuit} is similar.

From \cref{increment_circuit,decrement_circuit} we can see that each circuit for $N$ bit occupations requires $N$ NOTs (classically controlled), $4N-2$ CNOTs (including both control conditions), and $N-1$ Toffoli gates (controlled on $|0\rangle$ in the first control and $|1\rangle$ in the second control).
One such circuit will be required for each distinct mode acted upon by the creation and annihilation operators in the Hamiltonian term (monomial), so if there are $f$ creation and annihilation operators in the term then implementing $O_F$ requires at most
\begin{equation}
	(6N-3)f
\end{equation}
gates and $f$ ancilla qubits (the `multiplier' qubits).

As discussed above, all that is required to implement the $O_F$ oracle for a creation/annihilation operator monomial is a sequence of the occupation increment and decrement circuits given in \cref{increment_circuit,decrement_circuit}.
For the $O_H$ oracle for the monomial, we have to calculate the value of the matrix element between some input state Fock state encoded in qubits $|x\rangle$ and its image $|y\rangle$ under the action of the oracle, which we will have already computed using $O_F$.
The action of a single creation or annihilation operator acting on a bosonic mode with occupation $w$ is given by
\begin{equation}
	a^\dagger|w\rangle=\sqrt{w+1}|w+1\rangle,\quad a|w\rangle=\sqrt{w}|w-1\rangle,
\end{equation}
assuming $w+1$ is still below the occupation number cutoff.
Hence, the matrix element of a product of creation and annihilation operators is
\begin{equation}
\label{matrix_element}
	\sqrt{\left(w^{(y)}_{n_\text{out}}w^{(y)}_{n_\text{out}-1}\cdots w^{(y)}_{2}w^{(y)}_{1}\right)\left(w^{(x)}_{n_\text{in}}w^{(x)}_{n_\text{in}-1}\cdots w^{(x)}_{2}w^{(x)}_{1}\right)},
\end{equation}
where the $w^{(y)}_{i}$ are the occupations of the modes in the output state $|y\rangle$ whose occupations are incremented by the monomial, and the $w^{(x)}_{i}$ are the occupations of the modes in the input state $|x\rangle$ prior whose occupations are decremented by the monomial.
When all of the occupation number changes for single modes are plus one or minus one, the variables $w^{(x)}_{i}$ and $w^{(y)}_{i}$ in \eqref{matrix_element} can be read directly from $|x\rangle$ and $|y\rangle$.
When multiple particles are created in the same mode, the occupations appearing in the product should be those \emph{immediately after} the corresponding creation operator is applied, and when multiple particles are annihilated in the same mode, the occupations appearing in the product should be those \emph{immediately before} the corresponding annihilation operator is applied.
Thus in either of these cases, additional addition or subtraction circuits as in \cref{increment_circuit,decrement_circuit} will be required to compute the variables in \eqref{matrix_element}.
If any of the `multiplier' qubits computed by $O_F$ are $|0\rangle$, then the whole matrix element should be $0$.

To compute the value of the matrix element given by \eqref{matrix_element} once the variables have been determined, we require $n_\text{in}+n_\text{out}-1$ multiplications followed by a square-root operation.
Reversible circuits for multiplication are given in e.g. \cite{parent18a}, the simplest of which requires $4W^2-3W$ Toffoli gates and linear ancilla space, where $W$ is the number of qubits used to store the value of the matrix element in \eqref{matrix_element} prior to taking the square-root.
Hence to implement all $n_\text{in}+n_\text{out}-1$ of the multiplications requires
\begin{equation}
	(4W^2-3W)(n_\text{in}+n_\text{out}-1)
\end{equation}
Toffoli gates and
\begin{equation}
	W\big(n_\text{in}+n_\text{out}+O(1)\big)
\end{equation}
ancilla qubits, because the multiplications are not in place and thus the partial products are stored at each step.
Note that since we only use the matrix element to compute the phase to be applied to the state, as described in Section IV in the main text, the fact that the ancillas are not uncomputed during the multiplications is no problem because the entire matrix element will be uncomputed after controlling the phase.
A reversible circuit for square-root is given in \cite{sultana11a}, requiring $\frac{5}{2}W^2+7W-2$ elementary gates and $\frac{1}{4}W^2+3W-2$ ancilla qubits.
Hence the total gate cost of the matrix element oracle $O_H$ is
\begin{equation}
	O\big(fW^2\big)
\end{equation}
and the number of ancilla qubits required is
\begin{equation}
	O\big(fW+W^2\big),
\end{equation}
where $W$ is the number of bits in the matrix element and $f$ is the number of creation and annihilation operators in the Hamiltonian term.

\vspace{-0.15in}
\section[appendix]{Proof of Lemma~\ref{decomp_lemma_1}}
\label[appendix]{lemma_1_proof_app}
\vspace{-0.05in}

This section contains of a proof of the following result stated in the main text, which is based on the proof of Lemma 4.3 in \cite{berry14a}:

\textbf{Lemma~\ref{decomp_lemma_1}.}
\emph{
    Any one-sparse Hamiltonian $H^{(1)}$ may be expressed up to to $L$ bits per real and imaginary part of each entry as a linear combination of
    \begin{equation}
        4L=4\left\lceil\log_2\left(\frac{\sqrt{2}\,\|H^{(1)}\|_\text{max}}{\gamma}\right)\right\rceil
    \end{equation}
    one-sparse, self-inverse Hamiltonians $G_j$, where $\gamma$ is the resulting error in max-norm.
    The $O_F$ oracles for the $G_j$ are the same as the $O_F$ oracle for $H^{(1)}$, and the $O_H$ oracle for any $G_j$ may be computed using two queries to the $O_H$ oracle for $H^{(1)}$.
}
\begin{proof}

Let $\Lambda$ denote the least power of two that is greater than $\|H^{(1)}\|_\text{max}$, the maximum magnitude of any entry in $H^{(1)}$.
Define $C^{\text{Re},l}$ and $C^{\text{Im},l}$ as the matrices of $l$th bits of the real and imaginary parts of $H^{(1)}$ (each of which is assumed to begin with the bit corresponding to $\Lambda/2$), so
\begin{equation}
\label{approx_decomp_app}
    H^{(1)}\approx\sum_{l=1}^L\frac{\Lambda}{2^{l+1}}\left(C^{\text{Re},l}+iC^{\text{Im},l}\right),
\end{equation}
where the approximation is due to cutting the entries off at $L$ bits.
The error in max-norm in \eqref{approx_decomp_app} is upper-bounded by
\begin{equation}
\label{decomp_error_app}
    \gamma\le\sqrt{2}\frac{\Lambda}{2^{L+1}}\le\frac{\sqrt{2}\,\|H^{(1)}\|_\text{max}}{2^L},
\end{equation}
since upon truncating to $L$ bits both the real and imaginary part in any entry may incur an error at most $\frac{\Lambda}{2^{L+1}}$, giving a total error of at most $\sqrt{2}\frac{\Lambda}{2^{L+2}}$ in any entry, and $\Lambda\le2\|H^{(1)}\|_\text{max}$.

Let $j_i$ be the index of the single nonzero entry in row $i$ of $H^{(1)}$ (or if there is no nonzero entry in row $i$, let $j_i$ be whatever index is computed by applying the $O_F$ oracle for $H^{(1)}$).
In terms of this, let us entrywise define the operators $G^{\text{Re},l,\pm}$ as follows:
\begin{equation}
\label{G_entry_form_app}
\begin{split}
    G^{\text{Re},l,\pm}_{ij_i}&=
    \begin{cases}
        1\quad\text{if $C^{\text{Re},l}_{ij_i}=1$},\\
        -1\quad\text{if $C^{\text{Re},l}_{ij_i}=-1$},\\
        \pm1\quad\text{if $C^{\text{Re},l}_{ij_i}=0$}
    \end{cases}\\&=
    \begin{cases}
        1\quad\text{if $\text{Re}[H^{(1)}_{ij_i}]>0$ and bit $l$ of $\text{Re}[H^{(1)}_{ij_i}]$ is $1$},\\
        -1\quad\text{if $\text{Re}[H^{(1)}_{ij_i}]<0$ and bit $l$ of $\text{Re}[H^{(1)}_{ij_i}]$ is $1$},\\
        \pm1\quad\text{if bit $l$ of $\text{Re}[H^{(1)}_{ij_i}]$ is $0$},
    \end{cases}
\end{split}
\end{equation}
and let all other entries in $G^{\text{Re},l,\pm}$ be zero.
Let the operators $G^{\text{Im},l,\pm}$ be defined similarly in terms of the $C^{\text{Im},l}$ or the imaginary parts of the entries in $H^{(1)}$.
From these definitions it follows immediately that
\begin{equation}
    C^{\text{Re},l}=\frac{1}{2}\left(G^{\text{Re},l,+}+G^{\text{Re},l,-}\right)
\end{equation}
and
\begin{equation}
    C^{\text{Im},l}=\frac{1}{2}\left(G^{\text{Im},l,+}+G^{\text{Im},l,-}\right)
\end{equation}
for each $l$, so
\begin{equation}
\label{approx_decomp_2_app}
    H^{(1)}\approx\sum_{l=1}^L\frac{\Lambda}{2^{l+2}}\left(G^{\text{Re},l,+}+G^{\text{Re},l,-}+iG^{\text{Im},l,+}+iG^{\text{Im},l,-}\right),
\end{equation}
with the error still given by \eqref{decomp_error_app}. The operators $G^{\text{Re},l,\pm}$ and $G^{\text{Im},l,\pm}$ have nonzero entries exactly where $H^{(1)}$ does, so they are one-sparse, and all of their nonzero entries are $\pm1$.
Therefore, they are not only one-sparse, but self-inverse as well, so \eqref{approx_decomp_2_app} is our desired decomposition.

\eqref{approx_decomp_2_app} contains $4L$ terms, and the error in max-norm is upper bounded by \eqref{decomp_error_app}.
This means that to achieve some target error $\gamma$, we require $L=\left\lceil\log_2\left(\frac{\sqrt{2}\,\|H^{(1)}\|_\text{max}}{\gamma}\right)\right\rceil$ bits in the real and imaginary parts of each entry, so in terms of $\gamma$ the number of terms in the decomposition is
\begin{equation}
    4L=4\left\lceil\log_2\left(\frac{\sqrt{2}\,\|H^{(1)}\|_\text{max}}{\gamma}\right)\right\rceil,
\end{equation}
as claimed.

Lastly, by construction the nonzero entries in $G^{\text{Re},l,\pm}$ and $G^{\text{Im},l,\pm}$ for each $l$ appear in the same locations as the nonzero entries in $H^{(1)}$, so the $O_F$ oracles for each $G^{\text{Re},l,\pm}$ and $G^{\text{Im},l,\pm}$ are the same as the $O_F$ oracle for $H^{(1)}$.
To compute the $O_H$ oracle for a given $G^{\text{Re},l,\pm}$, we first query the $O_H$ oracle for the same entry in $H^{(1)}$, which returns the value $H^{(1)}_{ij}$ of that entry in an ancilla register.
Then we can evaluate the corresponding entry in $G^{\text{Re},l,\pm}$ using the formula in the second part of \eqref{G_entry_form_app}: this is a three-qubit operation in which a single qubit storing the sign of the entry may be flipped, controlled on both the qubit storing the sign $\text{Re}[H^{(1)}_{ij}]$ and the qubit storing bit $l$ of $\text{Re}[H^{(1)}_{ij_i}]$.
Finally, we uncompute $H^{(1)}_{ij}$ using a second query to the $O_H$ oracle for $H^{(1)}$.

\end{proof}

\end{document}